\documentclass[
12pt,
a4paper,
amsmath,
superscriptaddress,
floatfix,
notitlepage,
longbibliography
]{revtex4-1}

\usepackage{graphicx,amssymb,amsmath,amsfonts}
\usepackage{epsfig}
\usepackage{xcolor}
\usepackage[mathscr]{eucal}
\usepackage{comment}
\usepackage{bm}
\usepackage{siunitx}

\usepackage{xspace}
\newcommand{\latin}[1]{{\it #1}}
\newcommand{\ie}{\latin{i.e.}\@\xspace}
\newcommand{\eg}{\latin{e.g.}\@\xspace}
\newcommand{\cf}{\latin{cf.}\@\xspace}

\newcommand{\viz}{\latin{viz.}\@\xspace}

\newcommand{\fig}[1]{Fig.~\ref{#1}}
\newcommand{\figs}[1]{Figs.~\ref{#1}}
\newcommand{\sfig}[1]{Supplementary Fig.~\ref{#1}}
\newcommand{\Fig}[1]{Figure~\ref{#1}}
\newcommand{\Figs}[1]{Figures~\ref{#1}}
\newcommand{\sect}[1]{Section~\ref{#1}}

\newcommand{\eq}[1]{Eq.~(\ref{#1})}

\newcommand{\pore}{\mathrm{in-pore}}
\newcommand{\bulk}{\mathrm{bulk}}
\newcommand{\wall}{\mathrm{wall}}
\newcommand{\wallion}{\mathrm{wall-ion}}

\newcommand{\MC}{\mathrm{MC}}

\newcommand{\etapore}{\eta_\pore}
\newcommand{\etabulk}{\eta_\bulk}
\newcommand{\ilconc}{c_\mathrm{IL}}

\newcommand{\Eself}{E_\mathrm{self}}
\newcommand{\ICC}{ICC$^*$\@\xspace}

\newcommand{\width}{w}
\newcommand{\widthacc}{w_\mathrm{acc}}

\newcommand{\length}{l}
\newcommand{\bulkSize}{H}

\newcommand{\espresso}{\mbox{\textsf{ESPResSo}}\@\xspace}
\newcommand{\pppm}{P$^3$M\@\xspace}

\usepackage{xr}
\begin{document}

\title{The effect of finite pore length on ion structure and charging}

\author{Konrad Breitsprecher}
\author{Manuel Abele}
\affiliation{Institute for Computational Physics, Universit{\"a}t Stuttgart,  Allmandring 3, D-70569 Stuttgart, Germany}

\author{Svyatoslav Kondrat}
\affiliation{Institute for Computational Physics, Universit{\"a}t Stuttgart,  Allmandring 3, D-70569 Stuttgart, Germany}
\affiliation{Institute of Physical Chemistry, Polish Academy of Sciences, Kasprzaka 44/52, 01-224 Warsaw, Poland}

\author{Christian Holm}
\affiliation{Institute for Computational Physics, Universit{\"a}t Stuttgart,  Allmandring 3, D-70569 Stuttgart, Germany}

\begin{abstract}

Nanoporous supercapacitors play an important role in modern energy storage systems, and their modeling is essential to predict and optimize the charging behaviour. Two classes of models have been developed that consist of finite and infinitely long pores. Here, we show that although both types of models predict qualitatively consistent results, there are important differences emerging due to the finite pore length. In particular, we find that the ion density inside a finite pore is not constant but increases linearly from the pore entrance to the pore end, where the ions form a strongly layered structure. This hinders a direct quantitative comparison between the two models. In addition, we show that although the ion density between the electrodes changes appreciably with the applied potential, this change has a minor effect on charging. Our simulations also reveal a complex charging behaviour, which is adsorption-driven at high voltages, but it is dominated either by co-ion desorption or by adsorption of both types of ions at low voltages, depending on the ion concentration.

\end{abstract}

\maketitle

\section{Introduction}

Electrical double-layer capacitors, or supercapacitors, are an important player on the market of energy storage devices. Supercapacitors store energy by electrosorption of counter-charge into the porous electrodes and provide high power densities and cyclability, but the stored energies are relatively low~\cite{miller:sci:08, simon_gogotsi:acr:13}. To increase the energy storage, electrodes with \emph{subnanometer} pores are used, which show anomalously high capacitances (per surface area)~\cite{pinero:carbon:06, gogotsi:sci:06, gogotsi:08} and hence high stored energies~\cite{kondrat:ees:12}. This anomalous increase of capacitance is due to the emergence of a \emph{superionic state}~\cite{kondrat:jpcm:11, kondrat:pccp:11}, \ie, screening of the electrostatic interactions in narrow conducting pores~\cite{rochester:cpc:13, schmickler:ec:14, mohammadzadeh_schmickler:ea:2015:NanotubeScreening}. In this work we shall focus on electrodes with such narrow pores only.

Modelling plays a crucial role in understanding and predicting the properties of supercapacitors, such as capacitance, energy storage and charging times. There have been many models and methods developed, but they can be conventionally spit into two classes. One class consists of models that literary `mimic' a supercapacitor, \ie, they consider an ionic liquid confined between two electrodes with porous structures, either with slit-shaped~\cite{borodin:10, feng:jpcl:11, wu:qiao:jpcl:12, vatamanu:jpcl:energystorage:13, vatamanu:acsnano:15} or cylindrical~\cite{borodin:10, shim:10} pores, or even featuring a complex pore network~\cite{merlet:natmat:12, burt_salanne:jpcl:16:ElectrolyteEffect}. However, since (typically molecular dynamics) simulations of such models are computationally demanding, they are necessarily `scaled-down' as compared to their experimental counterparts. For instance, the pore length in a typical simulation is tens of nanometers at best, while in the experimental systems the pores can be micrometer long, as follows from the carbon particle sizes (see, \eg, Refs.~\cite{li:07:ShortPores, dyatkin_gogotsi:jps:16}, and Ref.~\footnotemark[1]); likewise, the size of the region between the electrodes is of the order of nanometers in simulations, but it is hundreds of micrometers or millimeters in real supercapacitors.

On the other hand, there is a class of simplified models that consider an ionic liquid confined to a \emph{single} pore, either slit-shaped~\cite{kondrat:jpcm:11, kondrat:pccp:11, kiyohara:jpcc:07, asaka:jcp:10, kiyohara:jcp:11, jiang:nanolett:11, jiang:nanoscale:14, dudka:jpcm:16} or cylindrical~\cite{kornyshev:fd:14, lee:prl:14, schmickler:ea:2015:harmonicOscillator, rochester:jpcc:16}. These models are likely to describe more closely the long pores of real porous electrodes~\footnote[1]{While the electrodes with long well-defined slit pores exist~\cite{yang:sci13:graphene, lukatskaya:sci:13}, extended long pores might not be present in some porous materials, such as CDCs, which consist of interconnected networks of different pores (see \eg Ref.~\cite{oschatz_etal:carbon:16}). However, although such a network can be described as a collection of shorter pores, the ionic liquid is nevertheless mainly present deeply in the porous carbons and far from the contact with the bulk electrolyte or pore closings. In this sense, infinitely extended pores can still be considered as good models for these porous materials too. We note, however, that such pore networks and surface roughness may lead to less structured ionic liquid configurations inside the pores, which can influence the charging behaviour.}, but their deficiency is that the effects related to the pore closing and opening are ignored and the charging dynamics are not straightforward to study. Often such simplified models can be treated analytically~\cite{kondrat:jpcm:11, kornyshev:fd:14, lee:prl:14, schmickler:ea:2015:harmonicOscillator, rochester:jpcc:16, dudka:jpcm:16}, but also Monte Carlo (MC) simulations~\cite{kiyohara:jpcc:07, asaka:jcp:10, kiyohara:jcp:11, kondrat:pccp:11, rochester:jpcc:16} and classical density functional theories~\cite{jiang:nanolett:11, jiang:nanoscale:14,wu_li:arpc:07:DFT} have been applied.

The main goal of this work is to connect these two types of models and to study the effects of finite and infinite pore lengths on the ion structure and charging. To this end, we take a variation of the model of a single slit nanopore of infinite length developed in Refs.~\cite{kondrat:jpcm:11, kondrat:pccp:11}. We shall also present a \emph{new model} for supercapacitors, which consists of two electrodes with slit nanopores; in contrast to other similar models~\cite{feng:jpcl:11, wu:qiao:jpcl:12, vatamanu:jpcl:energystorage:13, vatamanu:acsnano:15}, our model considers pores that are open on one side only (\cf \fig{fig:model}a), \ie, we take into account the pore closings explicitly. We shall first show how the superionic state emerges in this model (\sect{sec:mm}; in this section we also describe the details of both models and the methods used to study them). Then, we shall discuss how the parameters of the two models can be connected (\sect{sec:0V}) and how to adjust the simulations and the analysis to make a meaningful comparison (\sect{sec:md}). The results obtained by these two models are compared and discussed in \sect{sec:mcmd:cmp}. In \sect{sec:pore_walls} we discuss how the ion structure and charging behaviour depend on the pore wall-ion interactions. We summarize and conclude in \sect{sec:summary}.

\section{Models and  Methods}
\label{sec:mm}

\begin{figure}
\includegraphics[width=0.9\textwidth]{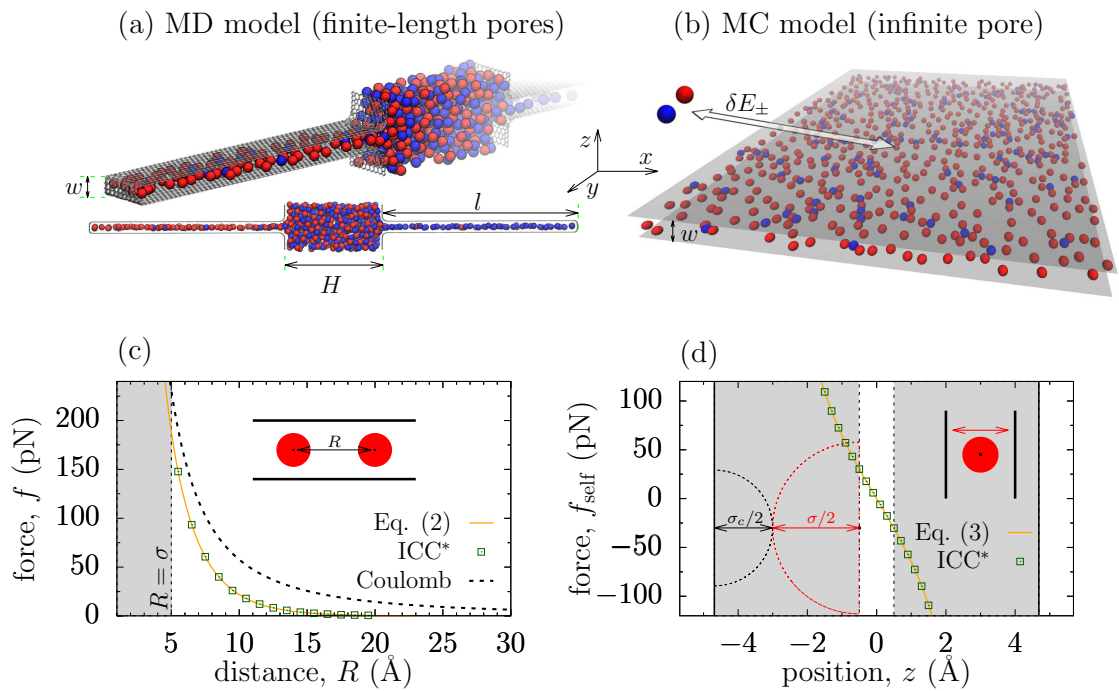}
\caption{Supercapacitor models in Monte Carlo (MC) and molecular dynamics (MD) simulations. (a) In MD simulations an ionic liquid is confined between two electrodes with a slit pore of \emph{finite} length each. The electrodes are constructed from the carbon atoms, and the electrostatic potential $2U$ is applied between the electrodes. (b) In Monte Carlo simulations the ions are confined in a single metallic slit pore, \emph{infinitely} extended in the lateral directions. The electrostatic potential $U$ is applied to the pore walls with respect to the bulk (\ie, unconfined) electrolyte. (c) Force $f$ between two ions confined in a metallic slit pore of width $\width=9.37$\AA. The open squares denote the force calculated by the \ICC method in the middle of a finite pore; this method is used for constant-potential MD simulations. The solid line shows the force obtained directly from the interaction potential for an \emph{infinite} pore, \eq{eq:mc:u2}, \ie, $f_{\alpha\beta} = -d v_{\alpha\beta}/dR$ (note that $f_{++} = f_{--} = - f_{+-} \equiv f$). The ion diameters are $\sigma = \sigma_\pm = 5$\AA~and their centers are located on the symmetry plane of the slit, \ie, $z_1 = z_2 = 0$ in \eq{eq:mc:u2}. The Coulomb force (dash line) is shown for comparison. (d) Force $f_\mathrm{self}$ acting on a single ion inside a slit metallic pore due to image forces. The open squares show $f_\mathrm{self}$ calculated in the middle of a finite pore by using the \ICC method, which is compared with the force for the infinitely long pore, $f_\mathrm{self} = -d\Eself/dz$ obtained from the interaction potential (\ref{eq:mc:u1}) used in the MC simulations (solid line). 
\label{fig:model}
}
\end{figure}

We compare two models of nanoporous supercapacitors. In one model we consider a supercapacitor that consists of two electrodes separated by a distance $\bulkSize$, with each electrode featuring a slit nanopore of the specified width $\width$ and \emph{finite} length $\length$ (\fig{fig:model}a). This model will be used in MD simulations as described below; we shall call it the MD model.

In addition, we consider a model consisting of a single nanopore, \ie, an ionic liquid confined between two parallel metal plates, which are \emph{infinitely} extended in the lateral directions (\fig{fig:model}b); this model will be treated by MC simulations and will be called the MC model henceforth. We apply a potential $U$ at the plates of the MC model (see \sect{sec:methods:mc}), which by symmetry corresponds to the applied potential $2U$ between the two electrodes of the MD model.

In both models and in all simulations, we shall use the same pore width $\width=9.37$\AA. We note that the pore width accessible to the ions is smaller, as will be discussed below.

\subsection{Ionic liquid}

In both MD and MC simulations, we have taken charged soft spheres to model an ionic liquid. The ion-ion soft potential was the repulsive-only Week-Chandler-Anderson (WCA) potential~\cite{wca:jpc:71}, $\phi_\mathrm{WCA}$. We recall that $\phi_\mathrm{WCA}$ is the standard 12-6 Lennard-Jones potential cut at the ion-ion separation $R_\mathrm{min} = 2^{1/6} \sigma$, where $\sigma$ is the ion diameter, and shifted such that $\phi_\mathrm{WCA}=0$ at $R=R_\mathrm{min}$; this is to ensure that the corresponding force is a continuous function of $R$. 

The following parameters have been used in all simulations: Interaction parameter $\epsilon = 1$kJ/mol and the ion diameter $\sigma = \sigma_\pm = 5$\AA. In the solvent-free case, these parameters give the pressure $1$atm for the ion volume fraction $0.134$ (molar concentration $0.6$M) and at temperature $T=400$K.

\subsection{Pore walls}
\label{sec:mm:pore_walls}

In the MD model the electrodes have been constructed from carbon atoms, which, however, we modeled as WCA particles with the following parameters: $\sigma_c = 3.37$\AA~ and $\epsilon_{c}=1$kJ/mol. The hexagonal structure of the pore walls was obtained by subdividing the surface into equilateral triangles and placing the atoms in their centers. A side length of $2.5$\AA~ provides an atom-atom bond length of $1.44$\AA, similar as in graphene. The pore entrance was curved with a radius of $4$\AA, and the pore closing was curved with a radius of $2$\AA. 

In the MC model we neglect the pore wall structure and consider flat soft walls instead. To this end we propose the 10-4 Lennard-Jones (LJ) interaction potential
\begin{align}
\label{eq:phi104}
	\phi_\wallion^\MC (z) = 2 \pi \epsilon_\wallion \sigma_\wallion^2 \rho_\wall
	\left[ \frac{2}{5}\left(\frac{\sigma_\wallion}{z-z_0}\right)^{10} - 
		\left(\frac{\sigma_\wallion}{z-z_0}\right)^4 \right]
\end{align}
where $\epsilon_\wallion$ and $\sigma_\wallion$ are the wall-ion energy and diameter parameters, $\rho_\wall$ is the \emph{two-dimensional} number density of carbon atoms, and $z_0$ is the location of the wall. This potential is obtained by integrating the LJ inter-particle interaction potential over a \emph{surface} of LJ particles, where the surface is infinitely extended in the $(x,y)$ directions.

In order to match the MC and MD models, we have fitted the interaction potential (\ref{eq:phi104}) to the \emph{averaged} potential that an ion experiences when approaching an atomistic wall. We emphasize that (i) this fit is difficult to do accurately in the whole range of the wall-ion distances, and (ii) the atomistic wall-ion potential is not homogeneous in the lateral directions (\sfig{si:fig:wall-ion-wca}). We will discuss this in \sect{sec:pore_walls}, where we also consider the case of hard pore walls with the accessible pore width $\widthacc=6$\AA~for comparison.

\subsection{Grand canonical Monte Carlo simulations}
\label{sec:methods:mc}

In the MC model a slit pore is \emph{infinitely} extended in the $(x,y)$ directions (\fig{fig:model}b), which is modeled by applying periodic boundary conditions in these directions. The electrostatic potential $U$ was applied to the pore walls, which amounts to setting the electro-chemical potential to $\mu_\pm = \pm eU + \delta E_\pm$, where  $e$ is the elementary charge and $\delta E_\pm$ is an energy of transfer of a $\pm$ ion from the bulk of a supercapacitor into the pore (assumed equal for anions and cations). Note that $\delta E$ does \emph{not} include the change in the ion's self-energy (\cf \eq{eq:mc:u1}), and that its typical values lie between $-5k_BT$ and $45k_BT$~\cite{lee:prx:16}.

The electrostatic interaction energy between two ions confined in a metal slit pore is~\cite{kondrat:jpcm:11}
\begin{align}
\label{eq:mc:u2}
v_{\alpha\beta}(z_1, z_2, R) = 
	\frac{4q_\alpha q_\beta }{\varepsilon \width} 
		\sum_{n=1}^{\infty} K_0(\pi n R /\width)
			\sin\boldsymbol{(}\pi n(z_1+1/2)/\width\boldsymbol{)}
			\sin\boldsymbol{(}\pi n(z_2+1/2)/\width\boldsymbol{)}
\end{align}
where $q_{\alpha}$ and $q_\beta$ are the ion charges ($=\pm e$ in this work), $R$ is the lateral distance between the ions, $z_1, z_2 \in [-\width/2, \width/2]$ are their positions across the pore, and $\varepsilon$ is the dielectric constant (taken $\varepsilon = 4$ in this work). 

An ion confined in a narrow conducting nanopore experiences an image-force attraction to the pore walls. For a slit metallic pore, infinitely extended in the lateral directions, this interaction energy can be calculated analytically~\cite{kondrat:jpcm:11}
\begin{align}
\label{eq:mc:u1}
\Eself^{(\alpha)} (z)  = - \frac{q_\alpha^2 }{\varepsilon\width} \int_0^\infty\left[\frac{1}{2} 
	- \frac{\sinh\boldsymbol(k(1/2-z/\width)\boldsymbol)\sinh\boldsymbol(k(1/2+z/\width)\boldsymbol)}
			{\sinh(k)}\right]dk,
\end{align}
where $z$ is the position across the pore. Note that $\Eself^{(\alpha)}$ does not depend on the ion charge but only on its valency (taken $1$ in this work); thus, we shall omit index $\alpha$ in $\Eself^{(\alpha)}$.

The interaction potentials (\ref{eq:mc:u2}) and (\ref{eq:mc:u1}) constitute the superionic state. They were implemented in towhee simulation package~\cite{towhee, kondrat:pccp:11} and grand canonical Monte Carlo simulations were performed using the Widom insertion-deletion move~\cite{widomMC:1963}, translational move, and the molecular-type swap move~\cite{kondrat:pccp:11}. We performed $5\times10^6$ equilibration runs and up to $10^7$ production runs at temperature $T=400$K.

\subsection{Molecular dynamics simulations}

MD simulations have been performed using \espresso simulation package~\cite{espresso, limbach:cpc:06:Espresso, arnold_holm:espresso:2013} with the velocity-Verlet algorithm for integration and a Langevin thermostat (at temperature $T=400 K$ and damping constant $\xi=10$ps$^{-1}$) to model a NVT ensemble. 

For constant potential simulations, we used the \ICC algorithm \cite{arnold_holm:cpc:02:MMM2D, kesselheim10a, arnold_holm:entropy:13:ICC} in combination with the 3D-periodic electrostatic solver \pppm \cite{ballenegger08a}. In the \ICC method, the induced \ICC charges are defined on a discretized closed surface, and their values are determined iteratively each simulation step. Since the standard \pppm solver does not take into account the applied potential between the electrodes, we have additionally superimposed the corresponding external electrostatic potential, which has been pre-calculated numerically by solving the Laplace equation with the appropriate boundary conditions; this has been done iteratively on a equidistant lattice using a seven-point-stencil relaxation algorithm \cite{press07a}.  

To test how the superionic state emerges within the \ICC approach, we have calculated the force between two ions in the pore middle, and compared it with the force obtained from \eq{eq:mc:u2} as $f_{\alpha\beta}= - dv_{\alpha\beta}/dR$ (note that $f_{++} = f_{--}=-f_{+-} \equiv f$). \Fig{fig:model}c demonstrates an excellent agreement between the two methods. We have also compared the force due to image forces acting between an ion and the pore walls obtained by the \ICC approach and from \eq{eq:mc:u1} as $f_\mathrm{self}=-\Eself/dz$; again the agreement is very good (\fig{fig:model}d). We note however that the corresponding \emph{potential} acquires an additional contribution due to periodicity (see \sfig{si:fig:Eself}), which can be corrected by considering larger systems. However, this shift in $\Eself$ does not influence the ion-pore walls \emph{forces}, as \fig{fig:model}d demonstrates, and therefore the results of the MD simulations (\ie, it only shifts the energy level, but the energy differences remain the same).

\section{Non-polarized nanopores}
\label{sec:0V}

\begin{figure}
\includegraphics[width=0.7\textwidth]{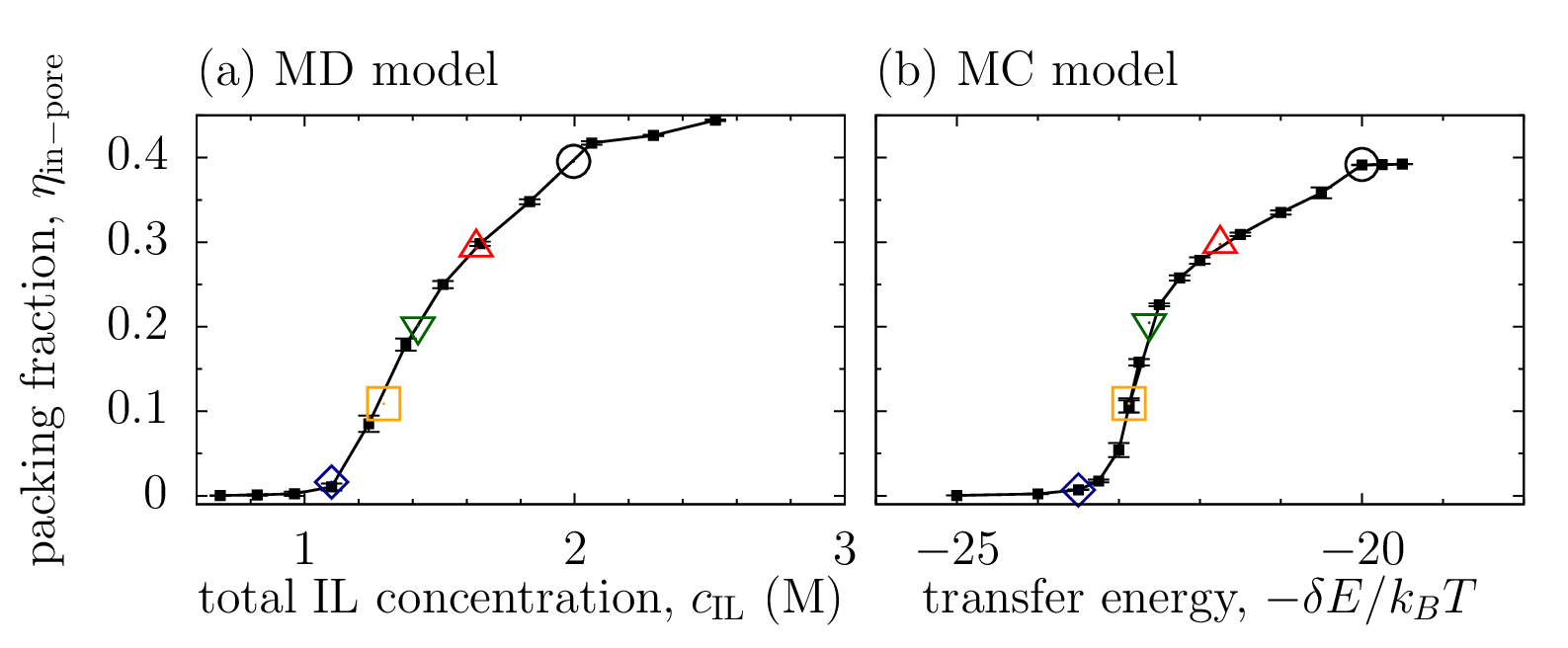}
\caption{Ion packing fraction in the pore at zero voltage from  molecular dynamics (MD) and Monte Carlo (MC)simulations. (a)~Ion packing fraction as a function of the total concentration $\ilconc$ of an ionic liquid (IL) from MD simulations. (b)~Ion packing fraction in the pore from MC simulations as a function of $\delta E$, an energy of transfer of an ion from the bulk electrolyte into the pore. Large symbols (rhombus, square, up and down triangles and circle) show the systems taken for a comparison of MC and MD models (\cf \figs{fig:mcmd:eta_pore} and \ref{fig:mcmd:cap}).
\label{fig:mcmd:eta}
}
\end{figure}

In order to facilitate a comparison between the two models (\fig{fig:model}a-b), we have matched the in-pore ion packing fractions at zero potential, $U=0$. We calculated this packing fraction as $\etapore = \pi \sigma^3/(6\Omega)$, where $\sigma=\sigma_\pm=5$\AA~is the ion diameter and $\Omega = S \widthacc$ is the volume of a pore; here $S$ is the lateral area and $\widthacc$ the accessible pore width. In the MD model, unless otherwise specified, only the middle parts of the pores were taken into account when calculating $\etapore$, \ie, the entrance and the closing of the pores were excluded (see \sect{sec:md:ion_struct}). Since $\widthacc$ is not known \textit{a priori} and is expected to vary with the applied potential, for definiteness we took $\widthacc = \width - \sigma_c$, where $\sigma_c = 3.37$\AA~is the diameter of the carbon atom. The pore width $\width=9.37$\AA~gives $\widthacc=6$\AA. Note that this is the accessible pore width for a system with hard pore walls.

In the MC model, the pore occupation is controlled by the ion transfer energy $\delta E_\pm = \delta E$. \Fig{fig:mcmd:eta}a shows that the in-pore packing fraction decreases as $\delta E$ increases, and the pore becomes more ionophobic~\cite{kondrat:nh:16, lee:prx:16}.

In the MD model, the in-pore ion packing fraction can be controlled by changing the total concentration of ions in a supercapacitor, $\ilconc$. Physically this can be realized by varying the pressure in the case of neat ionic liquids or by varying the salt concentration in the case of electrolyte solutions. \Fig{fig:mcmd:eta}b demonstrates that the pore becomes less populated as $\ilconc$ decreases. However, at extremely low concentrations our MD simulations predict the formation of ionic liquid clusters in the bulk electrolyte (\ie, between the electrodes), which influence the charging behaviour; we have therefore decided not to consider such cases in this work.

After having matched the pore occupancies at no applied potential, we perform voltage-dependent MC and MD simulations for the systems shown by symbols in \fig{fig:mcmd:eta}.

\section{Charging of nanopores of finite length (MD model)}
\label{sec:md}

There are two important parameters in the MD model that are not present in our MC system, \viz the pore length $\length$ and the size $\bulkSize$ of the bulk of a supercapacitor. In order to understand better their potential impact on charging, we first discuss how they influence the structure of an ionic liquid, in comparison to the MC model of a single infinitely-long nanopore.

\subsection{Ion structure}
\label{sec:md:ion_struct}

\begin{figure}
\includegraphics[width=0.45\textwidth]{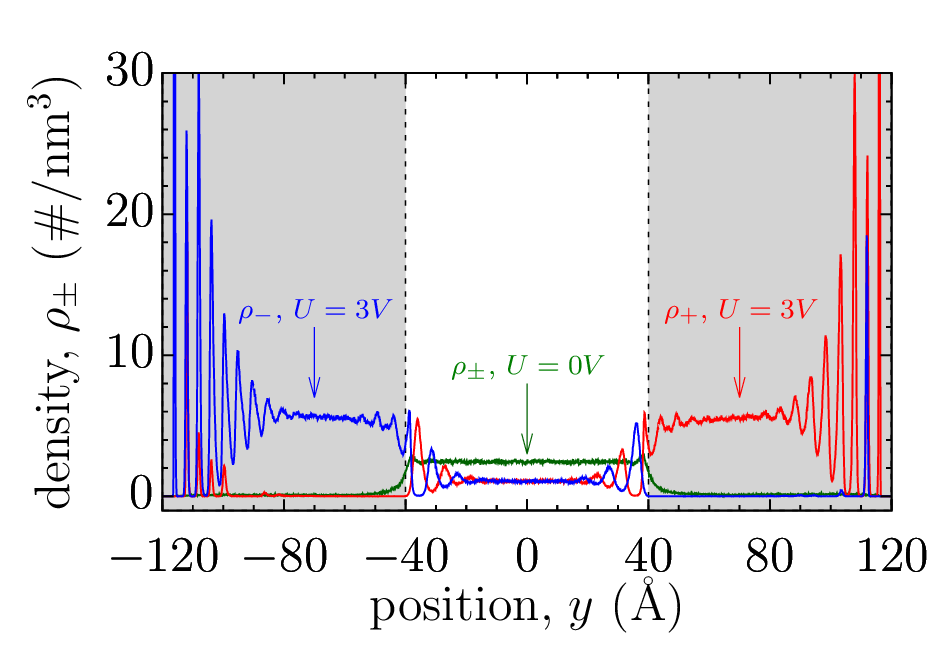}
\caption{Ion density profiles $\rho_\pm$ from MD simulations. The green line shows $\rho_+=\rho_-$ at no voltage, and the blue and red lines show $\rho_+$ and $\rho_-$ at applied potential $U=3V$. The grey areas denote the location of the pores. Without the appropriate correction steps, the co-ions get trapped in the pores, and the density of ions in the bulk electrolyte between the electrodes depends on the applied voltage. The total concentration of an ionic liquid in the supercapacitor is $\ilconc \approx 1.1$M.
\label{fig:md:rhoy}
}
\end{figure}

\Fig{fig:md:rhoy} shows the average ion density profiles $\rho_\pm (y)$ between the two electrodes for zero and non-zero applied potentials. We can make the following observations:
\begin{enumerate}

	\item The average ion density in the bulk electrolyte depends on the applied voltage. This implies that the chemical potential of the bulk ionic liquid changes with voltage, while it is taken constant in the MC model.

	\item Some \emph{co-ions} become trapped near the pore closings on the time scales of our MD simulations. This means that the system has not reached equilibrium. Note that this is unlikely to happen in the MC model as we perform grand canonical simulations.

	\item The ions exhibit a clear layering near the pore closings and openings, while they seem to form a nearly homogeneous structure in the middle of a pore. However, for non-zero potentials the counter-ion density \emph{is not constant along the pore} and increases from the pore entrance to the pore end. Clearly, in the MC model the average ion densities are position independent.

\end{enumerate}

We shall now discuss how to correct the MD simulations to be able to approach more closely the single-pore supercapacitor models. We will see, however, that although points (1) and (2) can be corrected relatively easily, point (3) is more subtle and makes it difficult to compare the MC and MD models \emph{quantitatively}.

\subsection{Bulk density calibration}
\label{sec:md:bulk}

\begin{figure}
\includegraphics[width=0.8\textwidth]{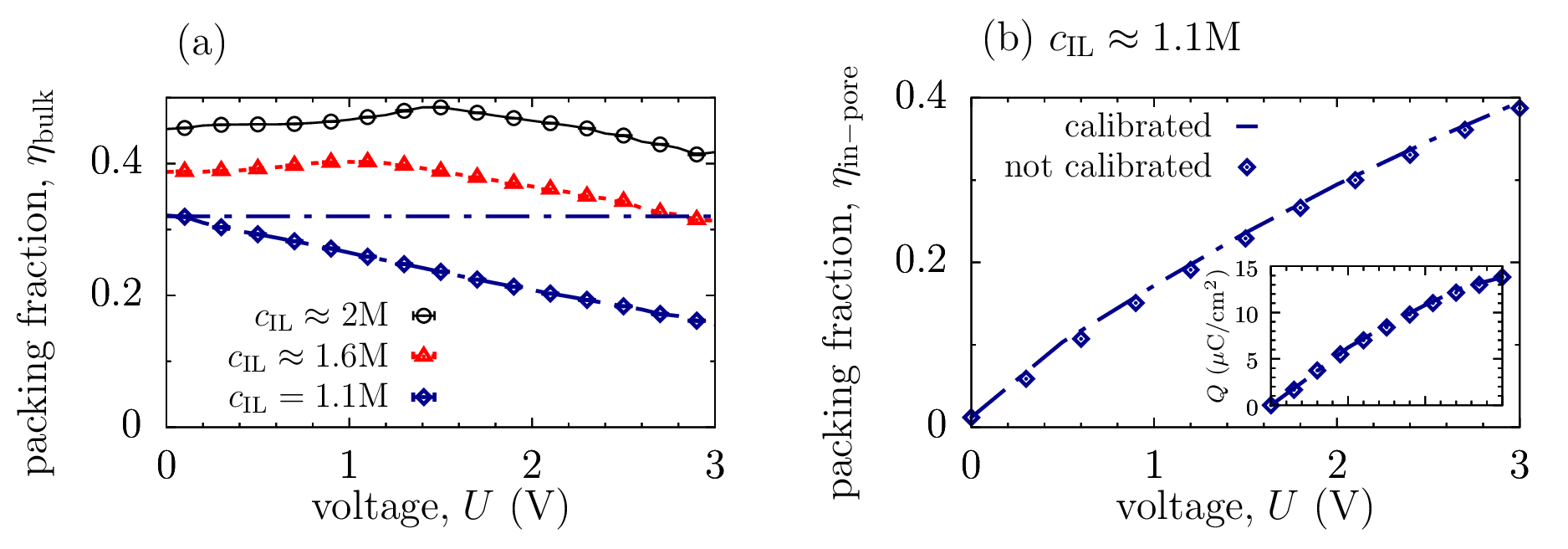}
\caption{Calibration of bulk density in MD simulations. (a) Total packing fraction $\etabulk$ of the ions in the \emph{bulk} electrolytes between the electrodes (see \fig{fig:model}a); $\etabulk$ can change significantly with the applied potential. To keep $\etabulk$ constant (dash-dot line for $\ilconc \approx 1.1$M) we calibrate the bulk density during equilibration runs by inserting ion pairs into the system (see the main text). (b) Total packing fraction $\etapore$ of the ions in the pore with and without density calibration for $\ilconc=1.1$M; for the remaining two concentrations from panel (a) see \sfig{si:fig:md:bulk_calibration}. Such a calibration influences $\etapore$, albeit weakly, but its effect on the accumulated charge is negligible, however (see the inset).
\label{fig:md:calibration}
}
\end{figure}

Clearly, also in experimental systems and commercially fabricated supercapacitors, the ion density in the bulk (\ie, between the supercapacitor electrodes) can vary with the applied voltage. However, the volume of a bulk region in these systems is typically large, as compared to the total pore volume, and this change is expected to be small. In MD simulations, the size of a bulk region ($\bulkSize$) is often comparable to the pore size~\cite{feng:jpcl:11, wu:qiao:jpcl:12, vatamanu:jpcl:energystorage:13, vatamanu:acsnano:15}, and its effect on the bulk density can therefore be significant (\fig{fig:md:calibration}a).

One way to deal with this problem is to consider systems with sufficiently large $\bulkSize$, which would, however, increase the computational costs accordingly. We have therefore taken a different route. We chose to \emph{calibrate} the total number of ions in a system during equilibration runs each time the voltage is changed. This was done by inserting ion pairs into the system (or removing them from the system when necessary) until the bulk density $\rho_\bulk (U\ne 0)$ equilibrates to $\rho_\bulk(U=0)$. After the calibration we run production runs as usual.

Surprisingly at first glance, we have found that although the total ion density in the pore is slightly altered by calibration, it has practically no effect on the charge storage (\fig{fig:md:calibration}b). This is likely because a change $\Delta \mu_\pm$ in the chemical potential due to the change in the ion density is small compared to what the system gains from the applied potential, \ie, $\Delta \mu_\pm \ll eU$. We could not accurately estimate $\Delta \mu_\pm$ for our system, but we expect it to be of the order of few $k_BT$'s \cite{kato:jpcb:08:ILChemPot}. This is supported by an observation that the change in the transfer energy of about $3k_BT$ (in the MC model) corresponds to the change in the total ion concentration from $1$M to more than $2$M (see \fig{fig:mcmd:eta}). For comparison, $eU \approx 30 k_BT$ at $T=400$K and at an applied potential of $1$V.

\subsection{Avoiding ion trapping}

\begin{figure}[]
\includegraphics[width=0.8\textwidth]{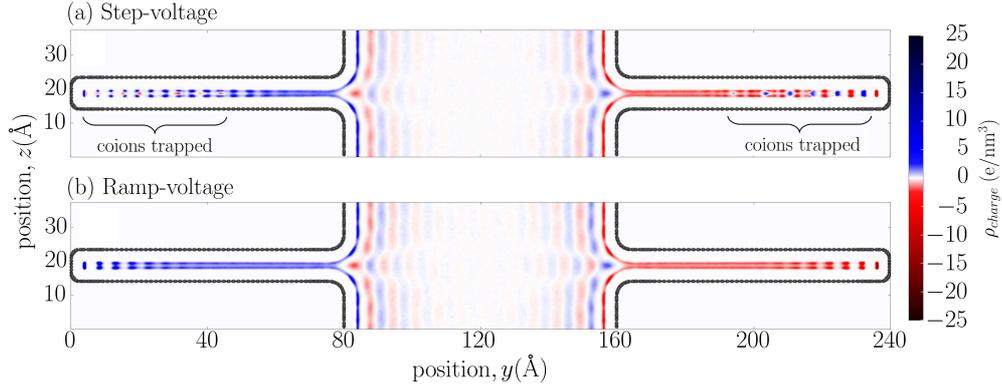}
\caption{Co-ion trapping in MD simulations. (a) Charge density heat-map for a step-voltage charging when a voltage of $2U=6$V is abruptly applied between the electrodes. Some co-ions are trapped near the pore closings even after $40$ns of simulation time. (b) Co-ion trapping can be avoided by charging the system slowly with a linear voltage ramp $U(t) = k t$, where $k=0.25$V/ns. There are no co-ions trapped in the pores after $10$ns of simulation time. These plots show the simulation results averaged over the last $8$ns of a $40$ns simulation.
\label{fig:md:trapping}
}
\end{figure}

\Figs{fig:md:rhoy} and \ref{fig:md:trapping}a show that at high voltages the co-ions in the pore become `trapped' near the pore closing on the time scales of our MD simulations ($\sim 100$ns). Such a co-ion trapping leads to a decreased charge storage and sluggish dynamics~\cite{pak_hwang:jpcc:16:IonTrapping}. To avoid ion trapping in this work we have used a linear \emph{voltage ramp} $U(t) = k t$ to charge our system, instead of a typically used step-voltage charging as in \fig{fig:md:trapping}a. \Fig{fig:md:trapping}b demonstrates that this strategy allows us to avoid co-ion trapping on computationally accessible time scales. 

The analysis of this approach will be presented in detail elsewhere.

\subsection{Effect of finite pore length, and pore entrance and closing}
\label{sec:md:pore_length}

\begin{figure}
\includegraphics[width=0.9\textwidth]{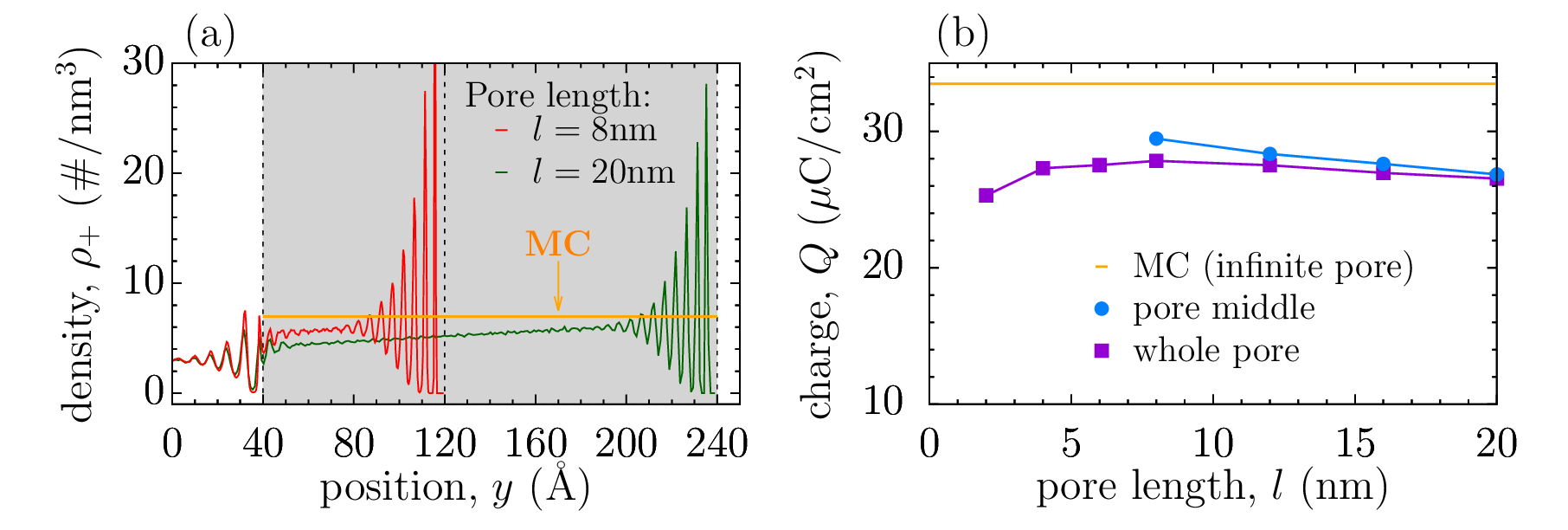}
\caption{Effect of pore length on ion structure and charging. (a) Counter-ion densities from MD simulations for different pore lengths $\length$ at applied potential $U=3$V (there are no coions in the pore at this voltage). The horizontal orange line shows the corresponding MC result. The MD results have been obtained using bulk calibration and ramp-voltage charging with the rate $k=0.25$V/ns. The grey areas highlight the location of the pores, and the vertical dash lines show the pore entrance and closings. Only half of a supercapacitor is shown. (b) Accumulated charge per surface area as a function of pore length. The charge has been calculated for the whole pore and when only the middle part of a pore is taken into account ($2$nm from the pore entrance and $3.5$nm from the pore end). The total concentration of an ionic liquid in the supercapacitor is $\ilconc \approx 1.6$M and the transfer energy in MC simulations is $\delta E = 21.75k_BT$. These parameters have been chosen such that the MC and MD models give the same ion densities at no applied voltage (see \fig{fig:mcmd:eta}).
\label{fig:mcmd:rhoy}
}
\end{figure}

We have pointed out in \sect{sec:md:ion_struct} that the pore entrance and closing influence the in-pore structure of ions, particularly strongly at non-zero voltages (\fig{fig:md:rhoy}). \Fig{fig:mcmd:rhoy}a shows that this behaviour remains true also for long pores ($\length=20$nm), \ie, the counter-ion density is systematically higher at the pore end and correspondingly lower where it begins; we note that these ion density profiles persist for long simulation runs (up to 200ns) and are likely the equilibrium profiles. Such a behaviour is understandable because image forces are weaker at the entrance, where the conducting pore walls end; in contrast, near the pore end the additional (closing) surface amplifies the image-force effects and hence attracts more ions. As a result, the counter-ion density along the pore changes approximately linearly between these two points.

Interestingly, the MC simulations predict the counter-ion densities that are closer to those at the pore end, rather than at the entrance (at the same applied potentials and for the same densities at zero potential). The exact reason for this is not clear to us, but it is tempting to speculate that it is actually the pore entrance that reduces the ion density, and in this way affects the ion concentration in the whole pore and hence its charging behaviour. On the other hand, it is also possible that the atomistic pore-wall structure in the MD model induces a weak ordering of an ionic liquid, and thus leads to an additional entropic cost for dense ion packing (recall that the pore walls in the MC model are flat). This implies that in the MD model a higher voltage is needed to induce the same counter-ion density as in the MC model, and this is indeed what we observe (\cf \fig{fig:mcmd:eta_pore}). It is worth noting that similar effects have been reported for ionic liquids at flat (non-porous) electrodes~\cite{breitsprecher:jpcc:15:ElectrodModels}.


This inhomogeneity of the ion distribution manifests itself in the accumulated charge ($Q_\mathrm{whole-pore}$), which depends sensitively on the pore length and shows a non-monotonic behaviour (squares in \fig{fig:mcmd:rhoy}b). Interestingly, $Q_\mathrm{whole-pore}$ decreases as $l$ increases (for long pores), which is because the ion density at the pore entrance becomes less affected by the pore closing, attending a lower value (\fig{fig:mcmd:rhoy}a). Since the effects due to the pore closing and opening weaken with increasing $l$, $Q_\mathrm{whole-pore}$ approaches $Q_\mathrm{pore-middle}$ as $l\to\infty$, where $Q_\mathrm{pore-middle}$ takes into account only the middle part of a pore. However, both $Q_\mathrm{pore-middle}$  and $Q_\mathrm{whole-pore}$ are smaller than $Q_\mathrm{MC}$ obtained within the MC model featuring an infinitely long pore. As discussed, this is likely due to the effect that the pore entrance has on the in-pore ion density (\fig{fig:mcmd:rhoy}a).

Thus, \fig{fig:mcmd:rhoy} demonstrates that \emph{there is no well-defined bulk region inside the pore}, where the average ion density would be constant along the pore. This behaviour affects the charging behaviour and hinders a direct \emph{quantitative} comparison between the MC and MD models. We therefore restrict further discussions mainly to qualitative comparisons.

\section{Comparison of the results for finite and infinite pores}
\label{sec:mcmd:cmp}


We are now in position to compare the simulation results of the MD and MC models with finite and infinitely long nanopores, respectively. In the MD simulations, we have taken a voltage ramp of $k=0.25$V/ns whenever necessary (at voltages $\lesssim 2$V we observe no trapping for the pore length considered); the size of a bulk region was $\bulkSize=80$\AA. The pore length was $\length=80$\AA, but only the middle region of size $25$\AA\; was used to analyse the results (this is to exclude the contribution from strong ionic-liquid layering at the pore ends and entrances).

Since in MD simulations we had to adjust the \ICC\; charges each simulation step to keep a constant potential on the electrode surfaces, these \emph{constant-potential} simulations are computationally demanding, and we have performed them only for a limited number of voltages. As a result, since the differential capacitance and the charging parameter $X_D$ (see below) require numerical differentiations with respect to voltage, we have calculated them with lower resolutions. This can be contrasted with our MC simulations, in which the metallic nature of the electrodes is taken into account via the interaction potentials (\ref{eq:mc:u2}) and (\ref{eq:mc:u1}). This allows to reduce the computational costs significantly (note that such analytical solutions exist only for a few simple geometries~\cite{rochester:13}).

\subsection{Pore filling and charging mechanisms}

\begin{figure}
\includegraphics[width=0.42\textwidth]{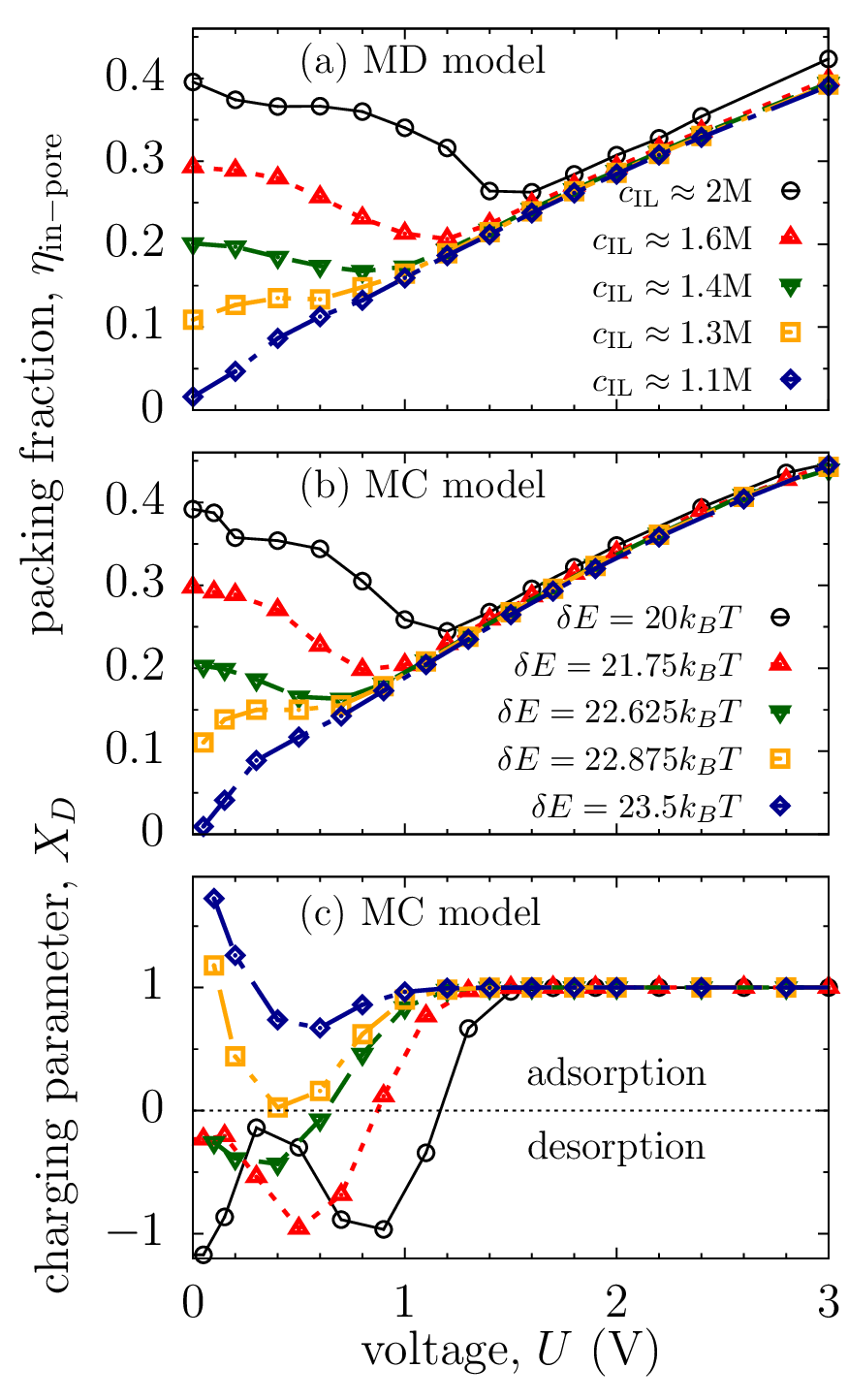}
\caption{
Total packing fraction $\etapore$ of ions in a pore and charging parameter $X_D$ from (a) MD and (b-c) MC simulations. Transfer energies ($\delta E$) and the ionic liquid concentrations ($\ilconc$) have been matched such as to have approximately the same  $\etapore$ at zero voltage. MC and MD models predict similar behaviours of $\etapore$ at low and intermediate voltages. Small discrepancies appear only at higher potentials likely due to the differences in the pore wall structures in the MD and MC models. (c) Charging parameter $X_D$, \eq{eq:XD}, obtained from MC simulations shows the regions where charging is dominated by adsorption ($X_D>0$) and desorption ($X_D< 0$). $X_D=0$ corresponds to swapping of co-ions for counterions. The same symbols and line codes are used in (c) as in (b). The comparison of $X_D$ obtained within the MC and MD models is shown in \sfig{si:fig:mcmd:XD}.
\label{fig:mcmd:eta_pore}
}
\end{figure}

\Fig{fig:mcmd:eta_pore} shows the total ion packing fraction $\etapore$ as a function of the applied potential, and demonstrates that pore filling proceeds similarly in the MC and MD models. Interestingly, in all cases of \emph{strongly ionophilic} pores, \ie, the pores with a substantial amount of an ionic liquid at no voltage, $\etapore$ first decreases for increasing voltage, and starts to increase only when there are no co-ions left in the pore. In other words, at low voltages charging is dominated by \emph{co-ion desorption}, while it is the \emph{counter-ion adsorption} that drives charging at higher applied potentials~\cite{kondrat:jpcm:11, wu:qiao:jpcl:12, vatamanu:jpcl:energystorage:13, kondrat:nh:16}. This is in agreement with the recent observation~\cite{kondrat:nh:16} showing that desorption (and swapping) are thermodynamically preferable over adsorption in most cases, except of a narrow window of parameters in which desorption and swapping are infeasible due to the lack of co-ions.

	To characterize charging mechanisms in more detail, we introduce a charging parameter, similar to the parameter $X$ of \citeauthor{forse:jacs:16:chmec}~\cite{forse:jacs:16:chmec},
\begin{align}
\label{eq:XD}
X_D(U) = \frac{e}{C(U)}\frac{d N}{d U},
\end{align}
where $e$ is the elementary charge, $C(U)= dQ/dU$ the differential capacitance, $Q$ denotes the accumulated charge and $N$ the total number of ions. $X_D$ expresses how charging is related to pore filling or de-filling, and thus describes which charging mechanism takes place. If charging is driven solely by swapping of coions for counter-ions, then the total ion density does not change, $N=\mathrm{const}$, and hence $X_D = 0$. For pure electrosorption we have $edN/dU = dQ/dU$ and thus $X_D = 1$, while for desorption $dQ/dU = - edN/dU$ and so $X_D = -1$. The parameter $X$ of \citeauthor{forse:jacs:16:chmec} is related to $X_D$ in a similar fashion as the integral capacitance is related to the differential capacitance, \ie,
\begin{align}
	\label{eq:XI}
	X(U) = \frac{1}{Q} \int_{0}^{U} X_D (u) C (u)du,
\end{align}
which can be seen as a voltage-averaged $X_D$ with the weight $C(u)$, where $Q = \int_{0}^{U} C(u) du$ is a normalization constant.

The charging parameter $X_D$ obtained from MC simulations is presented in \fig{fig:mcmd:eta_pore}c. It shows that at high voltages charging is solely due to counter-ion adsorption, \ie, $X_D \approx 1$, but at low voltages it can be either co-ion desorption or counter-ion adsorption, depending on the transfer energy $\delta E$. Interestingly, for high values of $\delta E$, \ie, when the pore is nearly empty at no applied potential, the parameter $X_D$ is significantly greater unity, which means that \emph{both} counter and co-ions are adsorbed into the pore at low voltages. This is likely because the additional `ion pairs' screen the interactions between the counter-ions, reducing  the thermodynamic cost of adsorption (note that at low densities the entropic cost of ion insertion is low).

\subsection{Charging and differential capacitance}

\begin{figure}
\includegraphics[width=0.42\textwidth]{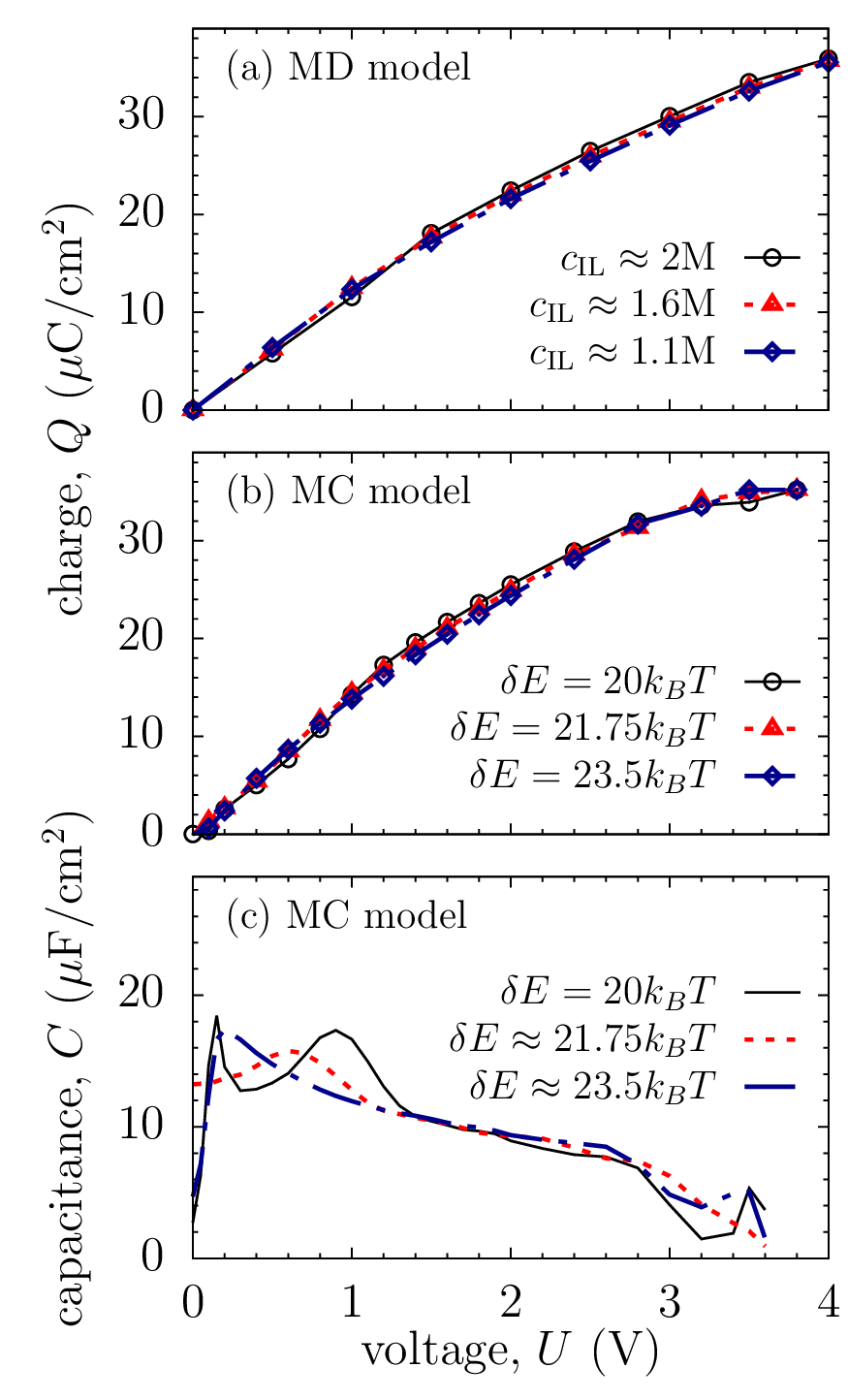}
\caption{Charging from MC and MD simulations. (a) Accumulated charge $Q$ as a function of applied potential $U$ from MD simulations for a few values of the total ionic liquid concentration $\ilconc$ in a supercapacitor. (b) $Q$ from MC simulations for a few values of the transfer energy $\delta E$. We have adjusted $\delta E$ and $\ilconc$ to give approximately the same in-pore ion packing fractions at zero voltage (see \fig{fig:mcmd:eta}). For a detailed comparison of the $Q(V)$ curves see \fig{fig:mcmd:walls}b and \sfig{si:fig:mcmd:charge}. (c) Differential capacitance $C$ as a function of voltage  from MC simulations. The comparison of the capacitances obtained within the MC and MD models is shown in \sfig{si:fig:mcmd:cap}.
\label{fig:mcmd:cap}
}
\end{figure}

Figure~\ref{fig:mcmd:cap} compares the accumulate charge $Q(U)$ from MC and MD simulations for pores with different occupancies at zero voltage, and demonstrates that also the charging process proceeds similarly in the MC and MD models. Interestingly, $Q(U)$ is practically independent of the transfer energies $\delta E$ (MC simulations) and ionic liquid concentrations $\ilconc$ (MD simulations). This is because the electrostatic contribution ($\pm eU$) to the total electrochemical potential dominates the contribution due to $\delta E$ and $\ilconc$, respectively (see \sect{sec:md:bulk}).

Fine details of the charging process are captured by the differential capacitance $C=dQ/dU$. Although $Q(U)$ does not seem to vary significantly with $\delta E$ (\fig{fig:mcmd:cap}b), $C(U)$ shows nevertheless a complex behaviour, particularly for densely populated ionophilic pores. For such pores, the capacitance exhibits a first maximum corresponding to the co-ion/counter-ion swapping and a second maximum associated with the co-ion desorption, before it finally decreases as the pore becomes more and more occupied by counter-ions at high voltages. For weakly ionophobic pores there is only one maximum in $C(U)$ at low pore occupancies, while at high potentials the charging proceeds similarly for all pores. For strongly ionophobic pores, the charging curves are shifted to higher voltages (\sfig{si:fig:mc:capen}). However, we have not been able to obtain such ionophobic pores in the MD model by varying the ionic liquid concentration $\ilconc$; thus, we shall not discuss this case further in this work.

Although charging in the MC and MD models show the same qualitative behaviour, there are some quantitative differences (\fig{fig:mcmd:cap}a-b, and \sfig{si:fig:mcmd:charge} and \ref{si:fig:mcmd:cap}), as discussed in \sect{sec:md:pore_length}.


\subsection{Ion structure}

\begin{figure}
\includegraphics[width=0.75\textwidth]{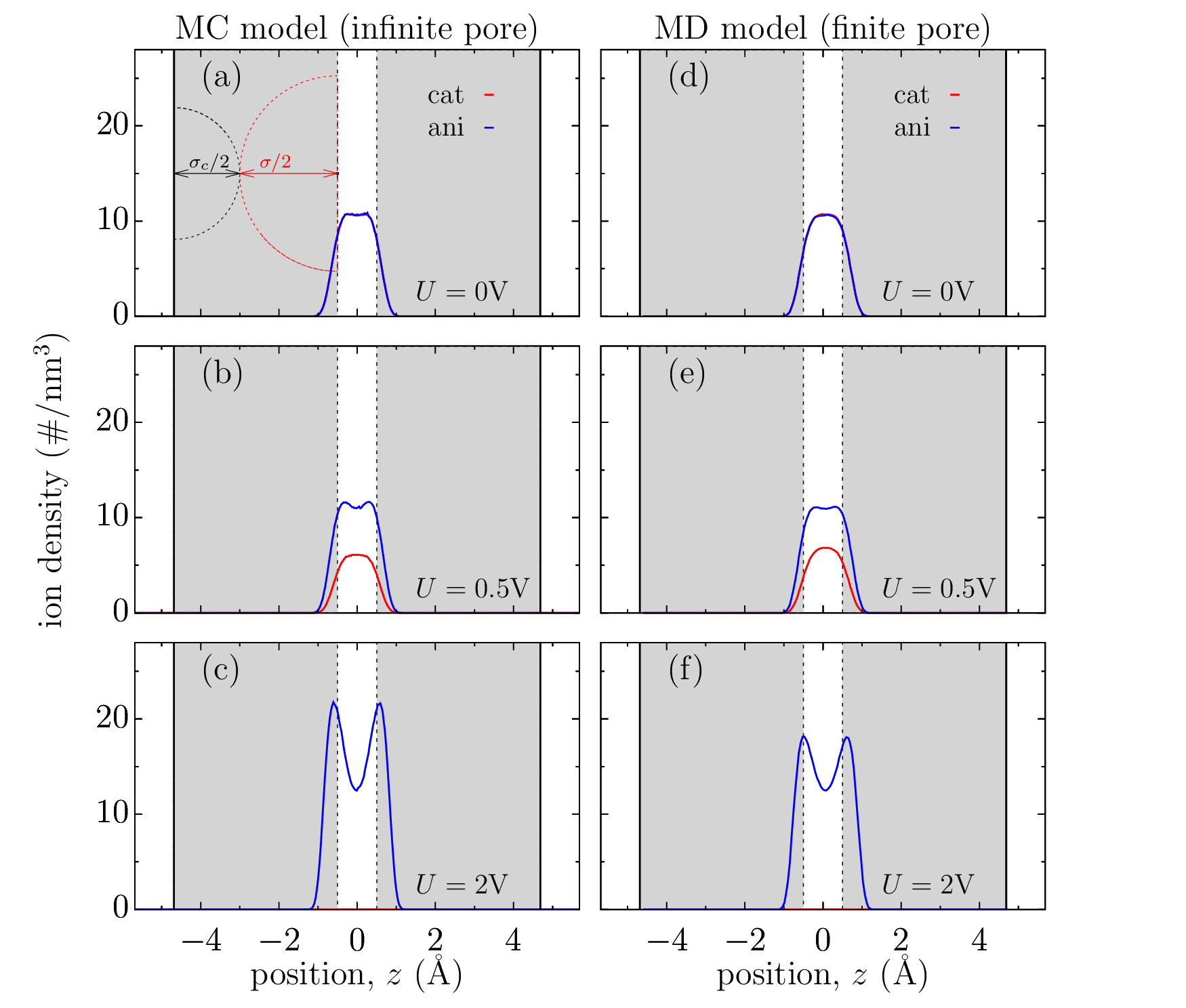}
\caption{
Ion packing in narrow pores from (a-c) MC and (d-f) MD simulations. The transfer energy in (a-c) is $\delta E_\pm = 21.75k_BT$. In (d-f) the molar concentration of an ionic liquid is $\ilconc \approx 1.635$M. The remaining parameters are the same as in \figs{fig:mcmd:eta_pore} and \ref{fig:mcmd:cap}.
\label{fig:mcmd:rhoz}
}
\end{figure}

We have also looked at the ion structure across the pore. This is shown in \fig{fig:mcmd:rhoz}, where we compare the density profiles obtained from MC and MD simulations for different voltages. The agreement is very good; small discrepancies are because we could not match exactly the ion densities at zero voltage.

At zero voltage the ion density has a maximum at the middle of the pore. This might seem surprising at first glance, since the image-force wall-ion attraction exhibits a \emph{maximum} at the pore center (\fig{fig:model}d). However, for ultranarrow pores considered in this work, this is altered by the wall-ion repulsive van-der Waals interactions, which produce a minimum rather than a maximum in the total wall-ion interaction potential (\sfig{si:fig:wall-ion}). Nevertheless, at high applied potentials, the counter-ions prefer to locate themselves at the pore walls. This is because the electrostatic energy ($\sim eU$) dominates the unfavourable van der Waals interactions between the walls and the ions, while the increased ion density pushes the counter-ions closer to the walls.

\section{Effect of pore walls}
\label{sec:pore_walls}

\begin{figure}
\includegraphics[width=.95\textwidth]{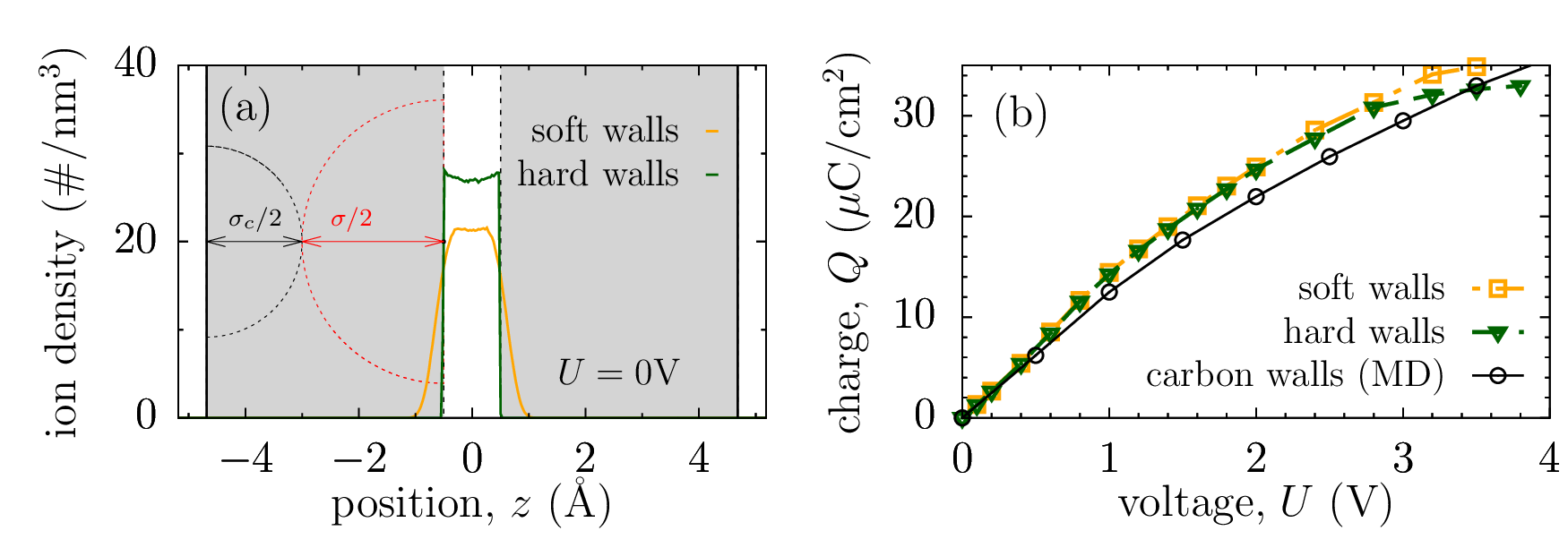}
\caption{
Effect of pore walls. (a) Ion density profiles across the pore in the case of the hard and soft pore walls at zero applied voltage. For the profiles at non-zero voltages see \sfig{si:fig:mc:rhoz}. (b) Accumulated charge as a function of applied voltage for pores with hard and soft walls obtained from MC simulations. The case of carbon walls (MD simulations) is shown for comparison. The ion transfer energies are $(\delta E)_\mathrm{hard} = 21.5k_BT$ and $(\delta E)_\mathrm{soft} = 21.75 k_BT$ for the hard and soft pore walls, respectively; their values have been chosen such that the ion packing fraction $\etapore \approx 0.3$ in the non-polarised pores.
\label{fig:mcmd:walls}
}
\end{figure}

In addition to the soft pore walls, which interact with the ions via \eq{eq:phi104}, we have considered the model of \emph{hard} walls, which are widely used in the literature~\cite{kondrat:pccp:11, kiyohara:jpcc:07, asaka:jcp:10, kiyohara:jcp:11, jiang:nanolett:11, jiang:nanoscale:14}. \Fig{fig:mcmd:walls} shows that the hard walls strongly influence the ionic liquid structure inside a pore, but their effect on the charge storage is moderate. For low voltages, the accumulated charge in both systems practically coincides, and the only significant differences arise at high applied potentials, where the pore with hard walls saturates while the soft-wall pore can accommodate more charge.

We have also considered the soft walls with the standard 9-3 Lennard-Jones interaction potential between the walls and the ions. This interaction potential is more difficult to fit to the average interaction potential between the ions and the atomistic wall of the MD model, which is not very surprising. We have found that the in-pore ion structure depends sensitively on the fitting parameters (results not shown). However, as in the case of the hard walls, it has no significant effect on the charging behaviour.

We can conclude that although fine details of the non-electrostatic wall-ion interactions are important for the ion structure, their impact on charging is minor, at least at low and intermediates voltages.


\section{Conclusions}
\label{sec:summary}

We have studied charge storage in supercapacitors with slit narrow pores using two models. In one model, treated by MD simulations, a supercapacitor consisted of two electrodes, each with a slit pore of a \emph{finite} length. In the second model, we focused on a single pore, \emph{infinitely} extended in the lateral directions; this model was studied grand-canonically by MC simulations (\fig{fig:model}). Our main conclusion is that although these two models are \emph{qualitatively} consistent with each other (\figs{fig:mcmd:eta_pore}, \ref{fig:mcmd:cap} and \ref{fig:mcmd:rhoz}), there are some important differences due to the finite pore length. In particular, the pore entrances and closings seem to have a vivid effect on the ion structure inside a pore. At high concentrations and/or high applied potentials, the ion density is not constant along the pore but varies roughly linearly between the pore entrance and the pore end, where it exhibit a strongly oscillatory structure (\fig{fig:mcmd:rhoy}). This impedes a direct \emph{quantitative} comparison of the two models.




We have also shown that:

\begin{itemize}

	\item In the MD model with finite pores, the ion density $\rho_\bulk$ between the electrodes of a supercapacitor can vary appreciably with the applied voltage. This can be corrected by calibrating $\rho_\bulk$ during equilibration runs, to keep it constant, as in the single-pore MC model. However, this change in the bulk density seems to have a minor effect on the charging behaviour (\fig{fig:md:calibration}). This result means that it is safe to consider relatively small electrode-electrode separations in supercapacitor models and refrain from computationally expansive bulk calibrations. Note that the studies of the charging dynamics would not be straightforward if the bulk calibration were necessary.


	\item At intermediate and high voltages (and for long pores), the co-ions become trapped in the pores on our typical simulation time sales ($\sim 100 $ns), producing non-equilibrium states. We have overcome this difficulty by using a voltage-ramp charging, instead of an abrupt step-voltage charging typically used in simulations (\fig{fig:md:trapping}). 

As co-ion trapping is expected to occur in experimental systems as well, we have studied this problem in more detail. In particular, we worked out a method to accelerate charging in systems with co-ion trapping and determined optimal charge/discharge regimes; the results of this study will be presented in a separate article.

	\item At high voltages, charging proceeds exclusively via counter-ion adsorption, while at low voltages the charging process is dominated by either co-ion desorption or counter-ion adsorption, depending on the ion transfer energy or the total ion concentration (\fig{fig:mcmd:eta_pore}). Remarkably, at high transfer energies, implying low ion concentrations, both counter and co-ions are adsorbed into the pore at low voltages (the charging parameter $X_D > 1$, see \eq{eq:XD} and \fig{fig:mcmd:eta_pore}c).

	\item Interestingly, the accumulated charge seems to be only weekly dependent on the total ions density in a supercapacitor (\fig{fig:mcmd:cap}). This observation provides an additional degree of freedom for optimizing the charging dynamics by varying the ion concentration without significantly compromising the energy density (note that the fine details of the charging process are resolved by the differential capacitance, which does depend on the total ion concentration/ion transfer energy, \fig{fig:mcmd:cap}c).

	\item Even though hard and soft pore walls lead to significant differences in the in-pore ion structure, they show practically the same charging behaviour (\fig{fig:mcmd:walls}). This is because the applied potential `overrules' all fine details of the non-electrostatic wall-ion interactions and the resulting ionic liquid structure.


\end{itemize}

In the context of the recent studies on ionophobicity of pores, it is instructive to emphasize that the pore occupancy at zero voltage, which determines the pore ionophobicity, can be varied by changing the total ion concentration ($\ilconc$) in a supercapacitor. However, by changing $\ilconc$ alone we could not achieve a state corresponding to strongly ionophobic pores, which provide high stored energies~\cite{kondrat:nh:16, lee:prx:16, lian_wu:jpcm:16:Ionophobic} and fast charging~\cite{kondrat:jpcc:13, kondrat:nm:14, lee:nanotech:14}. Thus, another method must be proposed to control effectively the ionophobicity of pores. 

Finally, we have considered only a few aspects of modeling supercapacitors and restricted our attention to charged soft spheres as a model for ionic liquids. Our results on the pore walls suggest that non-electrostatic interactions and image forces can have a profound impact on the in-pore ion structure. It will thus be interesting and fruitful to understand the effects due to the differences in ionic liquid models~\cite{breitsprecher14a:CG, breitsprecher14b:CG}, and whether such simple and computationally inexpensive models can capture the charging behaviour correctly.

\section{Supplementary Material}

Supplementary figures show the ion-wall repulsive interaction potential (\fig{si:fig:wall-ion-wca}); the ion's self-energy across the pore (\fig{si:fig:Eself}); the total wall-ion interaction potential (\fig{si:fig:wall-ion});  the effect of calibration on the ion density and accumulated charge (\fig{si:fig:md:bulk_calibration}); the comparison of the charging parameter $X_D$ (see \eq{eq:XD}) obtained within the MD and MC models (\fig{si:fig:mcmd:XD}); the accumulated charge within the MD and MC models (\fig{si:fig:mcmd:charge}); the differential capacitance within the MD and MC models (\fig{si:fig:mcmd:cap}); the differential capacitance and stored energy from MC simulations (\fig{si:fig:mc:capen}); and the ion density profiles across the pore for pores with soft and hard walls from MC simulations (\fig{si:fig:mc:rhoz}).

\begin{acknowledgments}
	C.H.~and K.B.~acknowledge the Deutsche Forschungsgemeinschaft (DFG) through the cluster of excellence ``Simulation Technology'' and the SFB 716 for financial support. We are also grateful for the computing resources on the Cray XC40 (Hazel Hen) from the HLRS in Stuttgart. This project has received funding from the European Unions Horizon 2020 research and innovation programme under the Marie Sk{\l}odowska-Curie grant agreement No 734276 (S.K.~contribution).
\end{acknowledgments}


%

\end{document}